\documentclass[%
reprint,
superscriptaddress,
amsmath,amssymb,
aps,
pra,
noeprint
]{revtex4-2}
\usepackage[utf8]{inputenc}

\usepackage{textcomp} 

\usepackage{graphicx}

\usepackage[capitalise]{cleveref}

\usepackage[T1]{fontenc}

\usepackage{siunitx}
\DeclareSIUnit{\amperehour}{Ah} 
\DeclareSIUnit{\hartree}{Ha}
\DeclareSIUnit{\molar}{\textsc{m}}
\DeclareSIUnit{\liter}{l}
\DeclareSIUnit{\rpm}{rpm}

\begin{document}

\title{Poly(1,4-anthraquinone) as an Organic Cathode Material: Simulation of Observable Bonding Properties to Li, Na, Mg, and Ca}

\author{Laura Femmer}
 \affiliation{Institute for Engineering Thermodynamics, German Aerospace Center (DLR), Wilhelm-Runge-Straße 10, 89081 Ulm, Germany}
 \affiliation{Helmholtz Institute Ulm (HIU), Helmholtzstraße 11, 89081 Ulm, Germany}

\author{Lukas Köbbing}
 \affiliation{Institute for Engineering Thermodynamics, German Aerospace Center (DLR), Wilhelm-Runge-Straße 10, 89081 Ulm, Germany}
 \affiliation{Helmholtz Institute Ulm (HIU), Helmholtzstraße 11, 89081 Ulm, Germany}

\author{Juliane Heitkämper}
 \affiliation{Institute for Engineering Thermodynamics, German Aerospace Center (DLR), Wilhelm-Runge-Straße 10, 89081 Ulm, Germany}
 \affiliation{Helmholtz Institute Ulm (HIU), Helmholtzstraße 11, 89081 Ulm, Germany}

\author{Sibylle Riedel}
 \affiliation{Helmholtz Institute Ulm (HIU), Helmholtzstraße 11, 89081 Ulm, Germany}

\author{Devran~Cay}
 \affiliation{Helmholtz Institute Ulm (HIU), Helmholtzstraße 11, 89081 Ulm, Germany}

\author{Florin Adler}
\affiliation{Institute of Organic Chemistry II and Advanced Materials, Ulm University, Albert-Einstein-Allee 11, 89081 Ulm, Germany}
\affiliation{CELEST Green Energy Lab Ulm, Ulm University, Lise-Meitner-Straße 16, 89081 Ulm, Germany}

\author{Birgit Esser}
\affiliation{Helmholtz Institute Ulm (HIU), Helmholtzstraße 11, 89081 Ulm, Germany}
\affiliation{Institute of Organic Chemistry II and Advanced Materials, Ulm University, Albert-Einstein-Allee 11, 89081 Ulm, Germany}
\affiliation{CELEST Green Energy Lab Ulm, Ulm University, Lise-Meitner-Straße 16, 89081 Ulm, Germany}

\author{Alexander J. C. Kuehne}
 \affiliation{Institute of Organic Chemistry III, Ulm University, Albert-Einstein-Allee 11, 89081 Ulm, Germany}
 \affiliation{CELEST Green Energy Lab Ulm, Ulm University, Lise-Meitner-Straße 16, 89081 Ulm, Germany}

\author{Zhirong~Zhao-Karger}
\email{zhirong.zhao-karger@kit.edu, birger.horstmann@dlr.de}
\affiliation{Helmholtz Institute Ulm (HIU), Helmholtzstraße 11, 89081 Ulm, Germany}
 \affiliation{Institute of Nanotechnology, Karlsruhe Institute of Technology (KIT), Hermann-von-Helmholtz-Platz 1, 76344 Eggenstein-Leopoldshafen, Germany}
 
\author{Piotr de Silva}
 \affiliation{Department of Energy Conversion and Storage, Technical University of Denmark, Anker Engelunds Vej 301, DK-2800 Kongens Lyngby, Denmark}

\author{Juan Maria García-Lastra}
 \affiliation{Department of Energy Conversion and Storage, Technical University of Denmark, Anker Engelunds Vej 301, DK-2800 Kongens Lyngby, Denmark}
 
\author{Birger Horstmann}
 \email{zhirong.zhao-karger@kit.edu, birger.horstmann@dlr.de}
 \affiliation{Institute for Engineering Thermodynamics, German Aerospace Center (DLR), Wilhelm-Runge-Straße 10, 89081 Ulm, Germany}
 \affiliation{Helmholtz Institute Ulm (HIU), Helmholtzstraße 11, 89081 Ulm, Germany}
 \affiliation{Department of Physics, Ulm University, Albert-Einstein-Allee 11, 89081 Ulm, Germany}

\begin{abstract}
    Poly(1,4-anthraquinone) (P14AQ) has emerged as a promising cathode material, offering high capacity and good cycling stability, yet the atomic-scale mechanisms governing metal-ion binding and electrochemical behavior remain poorly understood. To address this, we investigate the binding mechanisms of Li, Na, Mg, and Ca to P14AQ using quantum mechanical methods, particularly DFT and DFTB. 
    A key challenge lies in the material's structural complexity: multiple conformers of P14AQ are energetically similar but kinetically isolated due to significant energy barriers. To account for this, we develop an automated method to generate all unique P14AQ conformers for a periodic polymer chain without rotational duplicates through an orientation labeling scheme. 
    For each conformer, we systematically place a metal atom adjacent to every oxygen site, enabling a complete exploration of binding configurations. We observe two structural motifs: a single metal-oxygen bond and coordination to two opposite oxygen atoms. While Li and Na exhibit continuous energy distributions, Mg and Ca show an energy gap between the two motifs, with a strong preference for the two-oxygen binding configuration. 
    Galvanostatic measurements support these findings by showing lower gravimetric capacities for Ca and Mg than for Li and Na. For Na, the different numbers of metal–oxygen bonds are reflected in the two voltage plateaus observed experimentally. Overall, the combined computational and experimental results explain the higher capacity of monovalent ions in P14AQ: divalent ions cannot bind efficiently to a single oxygen site due to unfavorable energetics, and the conformational distribution of the polymer chain prevents optimal coordination. 
\end{abstract}

\maketitle

\section{Introduction}

Over the last decades, there has been a rapidly increasing energy demand, which has to be addressed as sustainably as possible. Due to more available renewable energy sources and new applications in need of electricity, the necessity for more energy storage systems like batteries is high \cite{larcherGreenerMoreSustainable2015, greySustainabilitySituMonitoring2017}. The reserves of lithium and other materials, like cobalt, which are used in lithium-ion batteries, are limited. This has led to a growing demand for new battery systems using more abundantly available materials \cite{larcherGreenerMoreSustainable2015, greySustainabilitySituMonitoring2017}. 

Sodium, magnesium, and calcium are less limited in their natural occurrence than lithium but also have a higher redox potential, reducing the full-cell voltage \cite{liCalciumChemistryNew2025, walterChallengesBenefitsPostlithiumion2020}. Ca and Mg are divalent ions and could therefore offer an increased gravimetric capacity compared to Li and Na, when used as metal anode. Furthermore, Ca and Mg have a lower tendency towards dendrite formation, which would increase battery safety \cite{zhao-kargerCalciumtinAlloysAnodes2022, ponrouchMultivalentRechargeableBatteries2019}. 

Noting that the aforementioned metals provide more sustainable options than Li, in addition, more long-term available cathode materials are needed. Organic materials seem to provide a viable option here, due to a higher abundance of raw materials and better recyclability \cite{qinRecentAdvancesDeveloping2020, luProspectsOrganicElectrode2020, sheaOrganicElectrodeMaterials2020}. Moreover, organic materials, which are used as electrode material, are loosely packed with weak intermolecular forces, which results in better ion transport properties \cite{xiuCombiningQuinoneBasedCathode2021, adlerSynthesisPoly14anthraquinoneUsing2025b, wesslingHowOrganicBatteries2025}. Because of electron delocalization across the organic material, organic electrodes also have a superior charge redistribution throughout the structure, making it beneficial especially for divalent ions \cite{xiuCombiningQuinoneBasedCathode2021, adlerSynthesisPoly14anthraquinoneUsing2025b}. 

\begin{figure}
\centering 
  \includegraphics[width=0.3\linewidth]{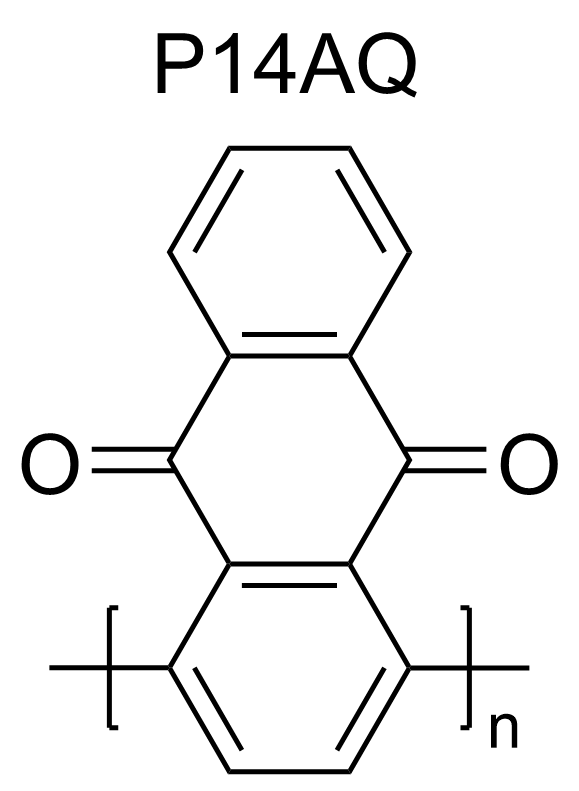}
  \caption{Poly(1,4-anthraquinone).}
  \label{fig:PAQ}
\end{figure}

Among the various organic materials, one possibility is poly(1,4-anthraquinone) (P14AQ), which is visualized in \cref{fig:PAQ}. It has a high theoretical capacity due to its two-electron redox behavior per subunit resulting from two carbonyl groups \cite{hauplerCarbonylsPowerfulOrganic2015}. Although the anthraquinone (AQ) units in the polymer are oriented nearly orthogonally, they exhibit significant electronic coupling arising from interactions between the carbonyl oxygen lone pairs and the $\pi$-system of neighboring units \cite{fornariUnexpectedlyLargeCouplings2019}. This lone-pair-$\pi$ conjugated structure enables intrachain charge delocalization while maintaining resistance to structural disorder over a range of oxidation states \cite{zhangPoly14anthraquinoneOrganicElectrode2025}. During cycling, the polymer undergoes only modest volume changes, enabling reversible electrochemical performance and good cycling stability \cite{zhangPoly14anthraquinoneOrganicElectrode2025}.

Compared to other quinone-based polymers, P14AQ shows a higher discharge capacity and cycling stability than poly(1,5-anthraquinone) (P15AQ) or poly(anthraquinonyl sulfide) (PAQS) with a Li anode \cite{songPolyanthraquinoneReliableOrganic2015}. Using Mg as the shuttle ion, P14AQ also shows a better cycling stability compared to poly(2,6-anthraquinone) (P26AQ) or PAQS \cite{panPolyanthraquinoneBasedOrganicCathode2016}. Besides Li and Mg anodes, a Ca anode was successfully tested with a P14AQ cathode \cite{zhao-kargerCalciumtinAlloysAnodes2022,riedelMechanicalEngineeringApproach2026}, and a Na anode was used with a thin film of anthraquinone and with an AQ powder \cite{wernerAnalysisOrderingEffects2021}.

In addition to experimental studies, quantum mechanical methods like density functional theory (DFT) \cite{zhangPoly14anthraquinoneOrganicElectrode2025, yangDensityFunctionalTheory2017, songPolyanthraquinoneReliableOrganic2015, rasheevFundamentalPromiseAnthraquinone2020, gallmetzerAnthraquinoneItsDerivatives2022} and density functional tight binding (DFTB) \cite{wernerAnalysisOrderingEffects2021, gallmetzerAnthraquinoneItsDerivatives2022} have been employed to study the atomistic properties of AQ. As a semi-empirical approximation to DFT, DFTB substantially reduces the computational cost at the expense of accuracy and is therefore especially attractive for large molecules \cite{elstnerDensityFunctionalTight2014, seifertCalculationsMoleculesClusters1996}.

Besides single AQ molecules, quantum mechanical computations of polymeric P14AQ chains have been shown in the literature, yet only a limited set of idealized structural configurations have been considered. These include alternating orientations (every second AQ unit adopts the same orientation) \cite{zhangPoly14anthraquinoneOrganicElectrode2025}, and antialternating arrangements (every second AQ unit adopts the opposite orientation) \cite{songPolyanthraquinoneReliableOrganic2015}. However, the full conformational space of P14AQ is far richer and the rotational freedom around the inter-unit bonds allows for a wide range of arrangements, including cases where only a single AQ unit could have a different orientation than all the others. To the best of our knowledge, a systematic exploration of the conformational landscape of P14AQ, particularly in relation to ion binding, has not yet been carried out.

To fill this gap, we develop an automated and computationally efficient method to systematically generate the conformational space of P14AQ, which we subsequently analyze using DFTB. Standard conformer search methods such as the AMS Conformer Generator \cite{rugerAMS2024101SCM2024} and CREST \cite{prachtAutomatedExplorationLowenergy2020} are primarily designed for isolated molecules and are not directly applicable to periodic polymer systems. Mapping the resulting non-periodic conformers onto a periodic unit cell generally leads to structural discontinuities at the cell boundaries. Furthermore, these approaches rely on stochastic conformational sampling. Consequently, they cannot guarantee complete and non-redundant coverage of the discrete conformational space defined by the periodic polymer. A tailored approach is therefore required that explicitly preserves the periodic connectivity while systematically generating all physically meaningful orientations of the AQ units.

To further investigate the binding behavior of P14AQ with Li, Na, Mg, and Ca, we systematically insert a single metal atom adjacent to each oxygen site across all conformers. The resulting structures are evaluated using both DFTB and DFT to assess stable geometries and energies. We compare the theory-based energy distributions with galvanostatic measurements of a battery cell with a respective metal anode and a P14AQ cathode. Our combined results reveal that experimentally observed capacities and voltage plateaus arise from distinct binding preferences: monovalent ions (Li, Na) can bind to multiple coordination sites across a broad range of conformers, while divalent ions (Mg, Ca) are restricted to specific configurations that satisfy two-site binding. This conformational dependence explains the lower capacity of divalent ions.

\section{Experimental methods}

To compare our theoretical insights with experimental data, we synthesize P14AQ, fabricate the P14AQ composite and corresponding electrodes, and assemble coin cells for galvanostatic charge–discharge testing. Furthermore, we determine the degree of polymerization of the synthesized P14AQ. The experimental procedures are discussed in this section.

\subsection{Synthesis of P14AQ}
We synthesize P14AQ from dichloroanthraquinone by dissolving it in dry dimethylformamide (DMF) and adding the solution to a mixture of $\text{Ni(COD)}_2$, cyclooctadiene (COD), and bipyridine (BPY) dissolved in DMF. Afterwards, we heat this mixture to $\SI{60}{\celsius}$ for $\SI{48}{\hour}$. After cooling down, we add hydrochloric acid and filter the precipitated solid. We perform sequential washes on the solid, followed by chloroform/methanol reprecipitation and suspension in dimethyl ether (DME). Finally, we dry the resulting bright yellow P14AQ powder in vacuum. More details about the synthesis are discussed in the Supporting Information in Section I \cite{yamamotoPolyanthraquinonesHavingPiConjugation1995, liStableEfficientElectrolytes2019a, zhao-kargerNewClassNoncorrosive2017, bulutNaBhfip4HfipOCHCF322011, ouldBoratesVsAluminates2024, riedelMechanicalEngineeringApproach2026}.

\subsection{Galvanostatic measurements}
\label{sec:methods_galvanostatic}
The synthesized P14AQ material is used to build two-electrode coin cells with different metal anodes and a P14AQ cathode. We then perform galvanostatic charge and discharge experiments with these cells. For the cell assembling, the synthesized P14AQ is prepared with Ketjenblack as a composite. This composite is mixed with sodium alginate, iso-propanol, and water and cast onto an aluminum current collector and dried. These cathodes contain approximately 60\% of P14AQ active material.

The P14AQ cathode is combined with a Li, Na, Mg, or Ca metal anode. We use metal foil for the Ca and Mg anodes, while pellets are used for the Li and Na anodes. We then punch both into circular discs. As electrolyte, we use the corresponding tetrakis(hexafluoroisopropyloxy)borate $\text{[B(hfip)}_4\text{]}$ salt dissolved in dry DME \cite{liStableEfficientElectrolytes2019a,zhao-kargerNewClassNoncorrosive2017,bulutNaBhfip4HfipOCHCF322011, ouldBoratesVsAluminates2024}.
We use a sandwich separator setup (CG / GF/C / CG). After assembling, all cells are rested for $\SI{1}{\hour}$. We give more details, especially about the electrolyte synthesis, in the Supporting Information in Section I \cite{yamamotoPolyanthraquinonesHavingPiConjugation1995, liStableEfficientElectrolytes2019a, zhao-kargerNewClassNoncorrosive2017, bulutNaBhfip4HfipOCHCF322011, ouldBoratesVsAluminates2024, riedelMechanicalEngineeringApproach2026}.

With these coin cells, we perform the galvanostatic charge and discharge experiments in a voltage range of $1.0\mbox{--}\SI{3.0}{\V} \text{ vs. Ca/Ca}^{2+}$, $0.5\mbox{--}\SI{2.5}{\V} \text{ vs. Mg/Mg}^{2+}$, $0.9\mbox{--}\SI{2.9}{\V} \text{ vs. Na/Na}^{+}$, and $1.1\mbox{--}\SI{3.1}{\V} \text{ vs. Li/Li}^{+}$ at a C-rate of $\SI{0.5}{\C}$ (P14AQ: $\SI{1}{\C}= \SI{260}{\milli \ampere \per \gram}$) using a Biologic BCS 805 testing unit. The results are discussed in Section \ref{sec:results_galvanostatic}.

\subsection{Degree of polymerization}
\label{sec:methods_dop}
The degree of polymerization, as a representation of the chain length of P14AQ, is crucial for deciding whether we perform molecular or periodic quantum mechanical calculations. Thus, we determine the degree of polymerization by performing polymer end-group analysis of the synthesized P14AQ and we cross check this value using gel permeation chromatography (GPC) and mass spectrometry. We briefly discuss the used methods here, more details are given in the Supporting Information in Section II \cite{grellChainGeometrySolution1998}.

We record $^1$H NMR spectra at $\SI{500.28}{\mega\hertz}$ in deuterated chloroform ($\text{CDCl}_3$). To record the spectra, we use a standard one-pulse sequence with 16 scans. All spectra are processed using standard procedures, including Fourier transformation, phase correction, and baseline correction.

For the GPC measurements, we dissolve samples of our synthesized P14AQ in $\text{CHCl}_3$, and filter them through PTFE filters before injection. We use an Agilent 1260 Infinity system equipped with refractive index (RI) and UV detectors. The separations are performed using linear/mixed-bed columns at a flow rate of $\SI{1.0}{\milli\liter\per\minute}$ and temperatures of $30\mbox{--}\SI{40}{\celsius}$. We calculate the number-average molecular weight, weight-average molecular weight, and dispersity relative to polystyrene standards, see Section \ref{sec:results_dop}. 

Moreover, we perform the mass spectrometry using a Bruker Flex Series MALDI-TOF mass spectrometer. We prepare samples of the synthesized P14AQ with the DCTB matrix and cationization agents. To confirm repeat unit masses and end groups, we use spectra in the laser desorption positive ion mode. Polymer distributions are evaluated acknowledging ionization bias toward lower molecular weights.

Evaluation of all these techniques, delivers a mean degree of polymerization of 66 (i.e. on average a polymer consists of this number of repeat units). The comparison of the results from the different techniques is discussed in Section \ref{sec:results_dop}. Due to the significant chain length estimate and to not consider chain-end effects, we apply periodic instead of molecular computational methods in the following.

\section{Theory-based methods}

Conformers are different spatial arrangements of the atoms in a molecule through rotation around single bonds \cite{rudinElementsPolymerScience2013}. To generate all conformers of P14AQ without rotational duplicates, we develop a computationally efficient approach. Representative examples of the resulting conformers are shown in \cref{fig:PAQ8_labels}. Subsequently, we automatically insert an individual metal atom into each of these conformers at chemically reasonable positions (specifically adjacent to an oxygen atom). The methodology is discussed in detail in this section. Further information on the quantum mechanical methods (DFT, DFTB, and force field) used for the investigation of the resulting structures can be found in the Supporting Information in Section III \cite{neeseORCAProgramSystem2012, neeseSoftwareUpdateORCA2022,grimmeR2SCAN3cSwissArmy2021,kruseGeometricalCorrectionInter2012,caldeweyherExtensionD3Dispersion2017,adamoReliableDensityFunctional1999,grimmeConsistentAccurateInitio2010, grimmeEffectDampingFunction2011a,weigendBalancedBasisSets2005,weigendAccurateCoulombfittingBasis2006,burschBestPracticeDFTProtocols2022,valeevLibintLibraryEvaluation2026,lehtolaRecentDevelopmentsLibxc2018, furnessAccurateNumericallyEfficient2020,rugerAMS2024101SCM2024,rappeUFFFullPeriodic1992,grimmeConsistentAccurateInitio2010,rugerAMSDFTB20241012024,grimmeConsistentAccurateInitio2010, grimmeEffectDampingFunction2011a,gausParametrizationBenchmarkDFTB32013, luParametrizationDFTB33OB2015, kubillusParameterizationDFTB3Method2015a,wernerAnalysisOrderingEffects2021, gallmetzerAnthraquinoneItsDerivatives2022,kresseInitioMolecularDynamics1993, kresseInitioMoleculardynamicsSimulation1994, kresseEfficientIterativeSchemes1996,hjorthlarsenAtomicSimulationEnvironment2017, bahnObjectorientedScriptingInterface2002,perdewGeneralizedGradientApproximation1996,grimmeConsistentAccurateInitio2010, grimmeEffectDampingFunction2011a,hortonAcceleratedDatadrivenMaterials2025, jainCommentaryMaterialsProject2013}. We perform all quantum mechanical calculations in vacuum.

\subsection{Rotation of one AQ unit}
\label{sec:methods_rotationAQ}
To highlight the importance of P14AQ conformers, we use molecular DFT calculations within ORCA to investigate the rotational freedom of the central AQ unit in a trimer chain. Two dihedral constraints are applied to enforce a planar structure (see \cref{fig:PAQ3_dihedral}), representing the extreme configuration of rotation around the inter-unit bond. We compare the energy of the constrained geometry with that of the unconstrained optimization. The result, which shows that the free rotation of an AQ unit through the plane of neighboring AQ units is prohibitive, is discussed in Section \ref{sec:results_conformers}. Since the AQ units are effectively locked in place, the consideration of distinct conformers is essential for a realistic description of the material's behavior.

\subsection{Conformer generation}
\label{sec:methods_conformers}

Given that rotations of the AQ units in P14AQ are energetically prohibitive, we automatically generate all relevant conformers under periodic boundary conditions. To avoid chain-end effects and to reflect the extended nature of P14AQ (estimated to consist approximately 66 AQ units per chain, as discussed in Section \ref{sec:results_dop}), we perform periodic quantum mechanical calculations.
To capture the full range of structural variability arising from different relative orientations of neighboring AQ units, we use unit cells with at least eight AQ units.

\begin{figure}
\centering
  \includegraphics[width=\linewidth]{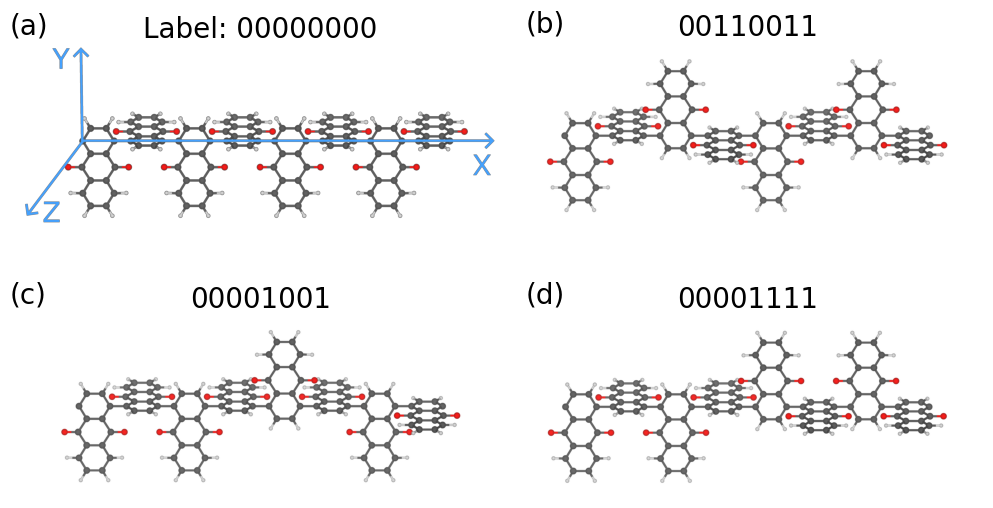}
   \caption{Four examples of unique conformers for 8 AQ units in the unit cell with the corresponding label. (a) The axes in the unit cell are shown. (a) and (b) depict the structural extreme cases. In (a) every second AQ unit has the same orientation, and in (b) every second AQ unit has the opposite orientation. These structures were visualized with VESTA \cite{mommaVESTA3Threedimensional2011}.}  
   \label{fig:PAQ8_labels}
\end{figure}

We assume each AQ unit is planar and that neighboring units are oriented at ${90^{\circ}}$ to one another. The polymer chain is aligned along the x-axis, as shown in \cref{fig:PAQ8_labels}a. Under these constraints, each AQ unit has two orientational options, defined by the direction of the oxygen-bearing ring.  We label these two options with 0 and 1. Note that it is sufficient to use two labels instead of four because the polymer is in a periodic unit cell and therefore the starting point is inconsequential. \cref{fig:PAQ8_labels} shows examples of the labels.

In the periodic unit cell, different symmetry operations give rise to equivalent representations of the same polymer with different labels. Structures that can be transformed into one another by translation of the unit cell along the x-axis, $90^\circ$ rotations around the x-axis, or $180^\circ$ rotations around the y- or z-axis belong to the same equivalence class. By identifying and eliminating such duplicates, we ensure a complete yet non-redundant set of conformers.

We automatically generate all binary sequences of length $N$, where $N$ is the number of AQ units in the unit cell, and identify duplicates by applying the symmetry operations above. For each equivalence group, we keep the representative with the numerically smallest label in the binary system. This approach is only possible for even $N$, as odd numbers would violate the $90^\circ$ angular constraint between neighboring AQ units at the unit cell boundary.

\begin{figure}
\centering
  \includegraphics[width=\linewidth]{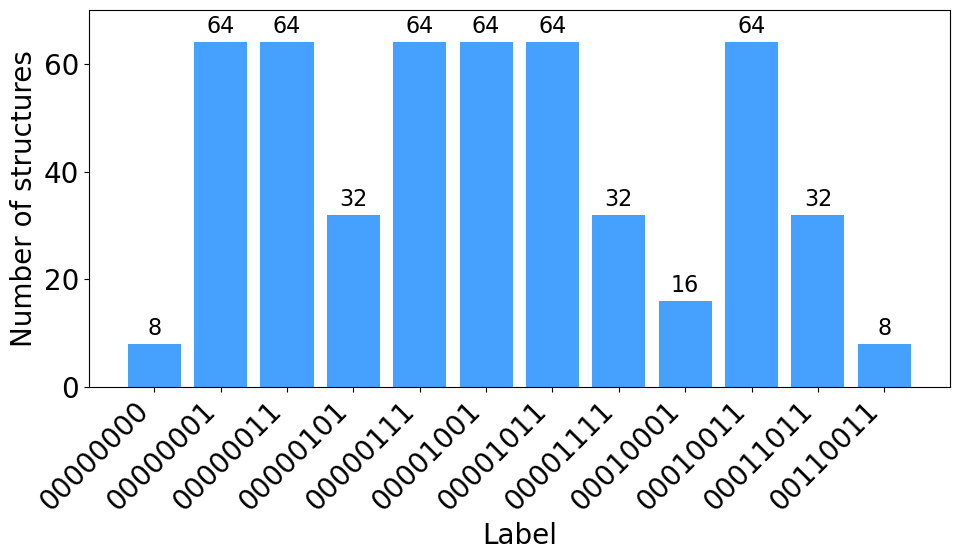}
  \caption{Number of structures in each equivalence class for 8 AQ units in the unit cell. The structures to the labels are shown in the Supporting Information in Figure S4.} 
  \label{fig:NrStructures_Label}
\end{figure}

\begin{figure}
\centering
  \includegraphics[width=0.7\linewidth]{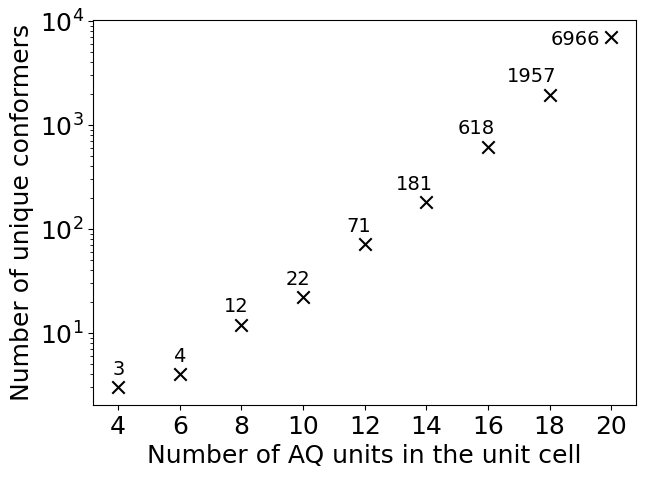}
  \caption{Number of unique conformers for different numbers of AQ units in the unit cell.}
  \label{fig:NrStructures_n}
\end{figure}

Considering 8 AQ units per unit cell yields 12 unique conformers. Their structures are depicted in the Supporting Information in Figure S4. \cref{fig:NrStructures_Label} shows the number of structures in each equivalence class. 
For a unit cell containing 12 AQ units, the number of unique conformers increases to 71. The number of resulting unique conformers for different numbers of AQ units in the unit cell is summarized in \cref{fig:NrStructures_n}. 

Following automated structure generation, we perform DFTB geometry optimizations to refine the structures and assess the energetic accessibility of all conformers, as discussed in Section \ref{sec:results_conformers}.

\begin{figure*}
\centering
  \includegraphics[width=0.9\linewidth]{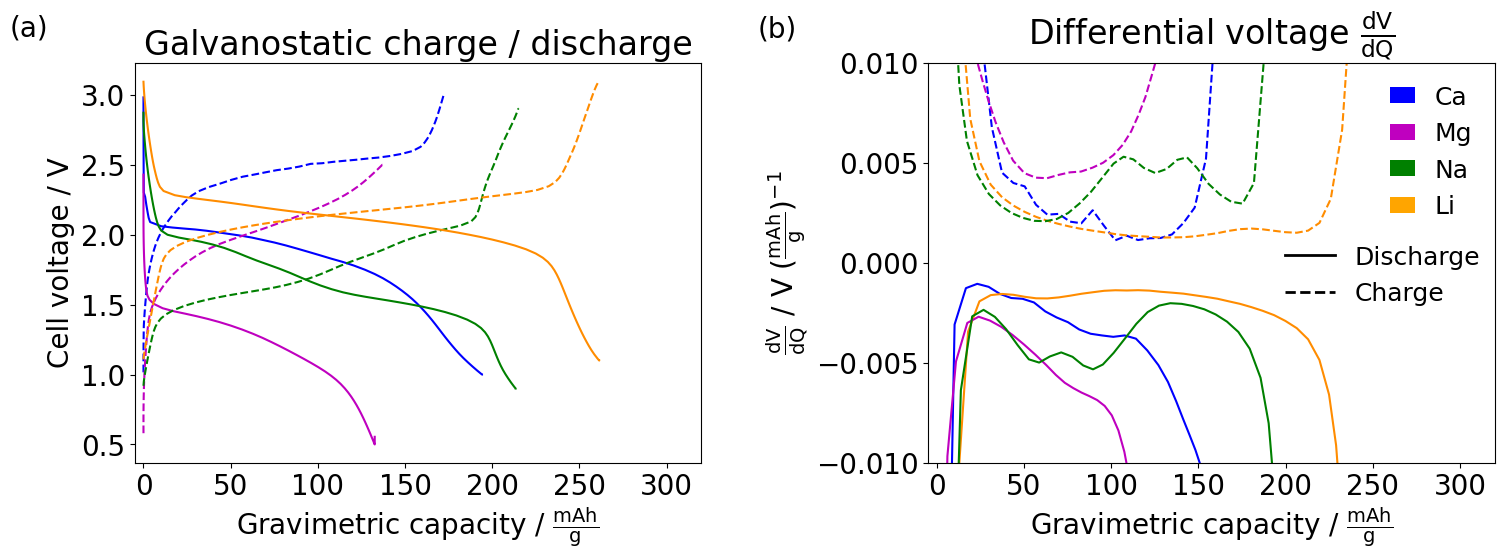}
  \caption{(a) Galvanostatic charge and discharge curves of Ca, Mg, Na, and Li anodes with a P14AQ cathode. The figure shows cycle two for Ca, Li, and Na, and cycle ten for Mg. We choose the cycle with a high gravimetric capacity and low overpotential. All measured cycles are shown in the Supporting Information in Figure S5. (b) Derivatives $\frac{\text{dV}}{\text{dQ}}$ of the charge and discharge curves in (a).}
  \label{fig:exp_data_onecycle}
\end{figure*}
\subsection{Insertion of individual metal atoms}
\label{sec:methods_metalinsertion}
To investigate the binding properties of different metal atoms to P14AQ, we automatically insert a single metal atom into each geometry-optimized structure individually, next to each oxygen atom. We insert the metal atom along the extension of the C=O bond, at a distance of $\SI{1.5}{\angstrom}$ from the oxygen atom. For conformers where the newly inserted metal atom coordinates are within $\SI{3}{\angstrom}$ of each other, we use the midpoint of the metal atom coordinates to reduce the number of resulting structures. 
Afterwards, we utilize pymatgen StructureMatcher \cite{ongPythonMaterialsGenomics2013} to remove duplicates by rotations. This approach results in 52 structures for the 8-AQ unit cells and 784 structures for the 12-AQ unit cells.

Due to the periodicity of the unit cell, it must be noted that the inserted metal atom is also repeated periodically. We discuss the influence of the periodically repeated metal atom in the Supporting Information in Section VI.

\section{Results}
In this section, we show the galvanostatic charge and discharge measurements of the battery cells with the P14AQ cathode and the different mono- and divalent metal anodes. Furthermore, we demonstrate why considering all conformers of P14AQ is necessary for a computational study. In the generated unique conformers, we insert a single Li, Na, Mg, or Ca atom and geometry-optimize the structures. Finally, we compare the resulting energy distributions to the experiments, unveiling the origin of the features observed in the experimental voltage curves. We refer to Ca and Mg as divalent atoms instead of ions because the calculations are performed charge-neutral.

\subsection{Galvanostatic charge and discharge experiments}
\label{sec:results_galvanostatic}

To investigate the performance of P14AQ cells with various metal anodes, we perform galvanostatic charge and discharge experiments with 0.5C as discussed in Section \ref{sec:methods_galvanostatic}. For this, we use two-electrode coin cells consisting of a P14AQ cathode, a metal anode, and a tetrakis(hexafluoroisopropyloxy)borate $\text{[B(hfip)}_4\text{]}$ based electrolyte. Regarding the metal anode and the electrolyte salt, we compare different metal elements, namely Li, Na, Ca, and Mg. 

\cref{fig:exp_data_onecycle}a shows the second charge and discharge cycle for the P14AQ cells with a Ca, Li, and Na anode and the tenth cycle for Mg. We choose the cycle with a high gravimetric capacity and low overpotential. All measured cycles are shown in the Supporting Information in Figure S5. Li shows the highest gravimetric capacity of $\SI{261}{\milli\amperehour \per \g}$ followed by Na with $\SI{214}{\milli\amperehour \per \g}$. 
Mg and Ca show smaller maximum capacities of $\SI{137}{\milli\amperehour \per \g}$ and $\SI{173}{\milli\amperehour \per \g}$, respectively, while charging. The maximum discharge capacity of Ca is higher, reaching $\SI{190}{\milli\amperehour \per \g}$, which could be the result of activation processes. The comparison of the monovalent to the divalent metal cells reveals that the cells with monovalent metal anodes reach a higher gravimetric energy density.

Additionally, the Na curves exhibit two voltage plateau regions with a small slope at around $\SI{1.5}{\V}$ and $\SI{2.0}{\V}$, as visible in \cref{fig:exp_data_onecycle}a. To further highlight these ranges, we show the derivative of voltage with respect to capacity, $\frac{\text{dV}}{\text{dQ}}$, of the charge and discharge curves in \cref{fig:exp_data_onecycle}b. The derivatives of the charge curves are positive, while the discharge derivatives are negative, because the charge curves strictly monotonically increase and the discharge curves strictly monotonically decrease in \cref{fig:exp_data_onecycle}a. 
The derivatives of the Na charge and discharge curves show two minima or maxima near zero, respectively. These points in the Na derivatives imply capacity ranges with a lower slope meaning voltage plateaus.

Similar to Na, the Li derivatives show two small peaks indicating two different phases. The absolute value of the derivative only increases very slightly between the peaks, resulting in a steadier increase or decrease in voltage for Li compared to Na. 

For Ca and Mg, one peak is visible with a previous or following increase in the absolute value of the derivative. This refers to a faster voltage drop. To elucidate the reason for the different obtained maximum capacities and the two voltage plateaus in the Na curve, we investigate the insertion of metals into P14AQ with our theory-based approach in Section \ref{sec:comparison_theoretical_experimental}.

\subsection{Chain length of P14AQ}
\label{sec:results_dop}
We measure the degree of polymerization, which determines the chain length of P14AQ, as described in Section \ref{sec:methods_dop}. Therefore, we perform polymer end-group analysis for the synthesized P14AQ and we obtain a mean degree of polymerization of 66. We cross-check this value using gel permeation chromatography (GPC) and mass spectrometry, where after correction for the different biases we obtain degrees of polymerization between 24 and 88, translating to number averaged molar mass $M_n = \SI{5000}{\g\per\mol}$ and weight averaged molar mass $M_w = \SI{18000}{\g\per\mol}$. Considering the large dispersity of $\text{\DJ}= 2.65$ (determined via GPC) and the consistent spacing of the polymers representing the repeat unit mass spectrometry, we conclude that the end-group analysis correlates well with the GPC and mass spectrometry results and we continue with the averaged degree of polymerization of 66.

\subsection{Conformers of P14AQ}
\label{sec:results_conformers}
\begin{figure}
\centering
  \includegraphics[width=0.4\linewidth]{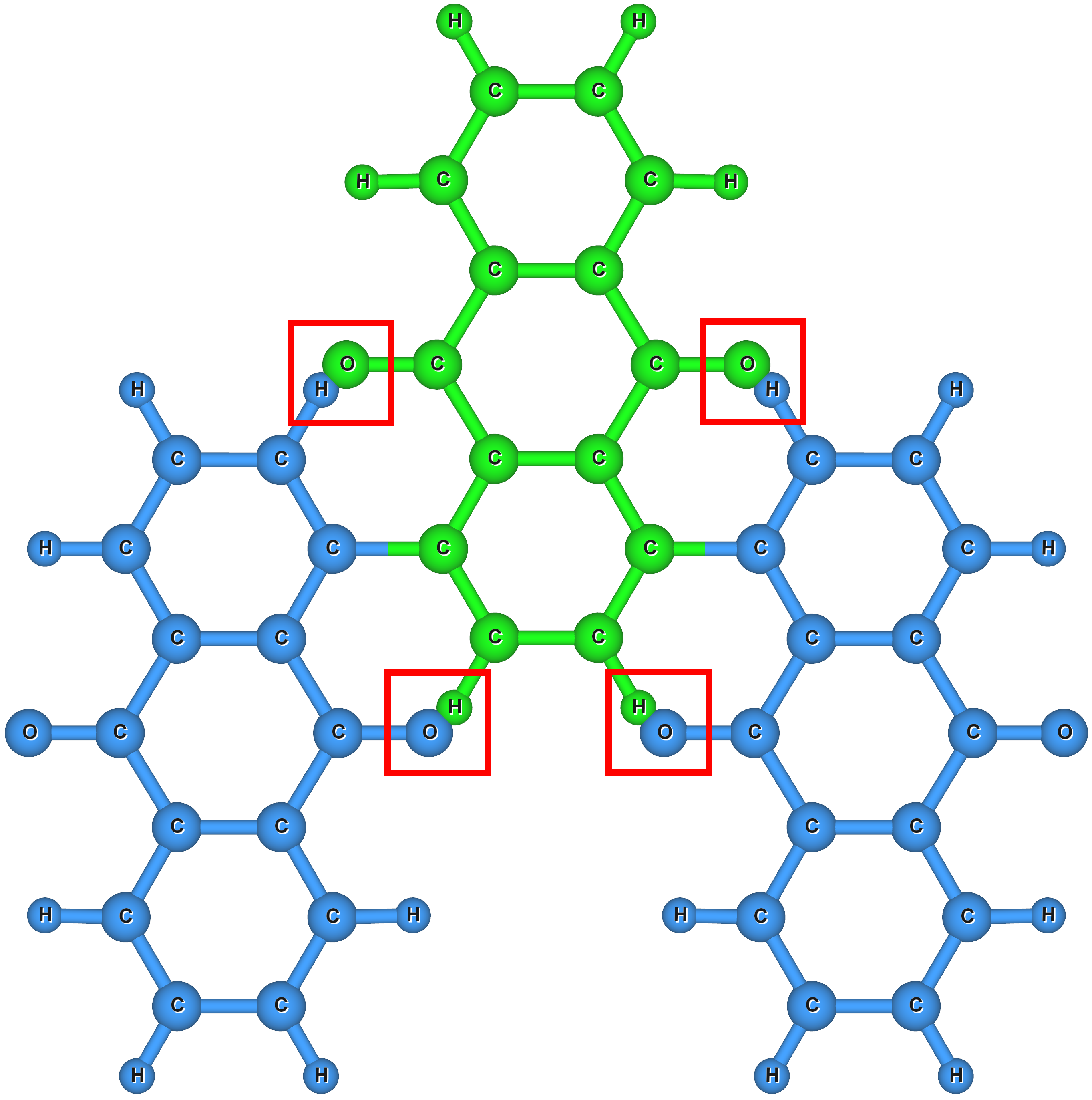}
  \caption{P14AQ with three AQ units in the same plane to illustrate the steric problems of a rotation of one AQ unit. The rotating AQ unit at the extreme position is colored in green. The red boxes indicate the steric problems between the hydrogen and oxygen atoms of the neighboring AQ units. For this visualization we utilized VESTA \cite{mommaVESTA3Threedimensional2011}.} 
  \label{fig:PAQ3_planar}
\end{figure}

To confirm the importance of P14AQ conformers, we investigate the feasibility of rotating one AQ unit relative to its neighbors. During such a rotation, the unit must pass through the plane of the adjacent AQ units, resulting in close spatial proximity between the oxygen atoms of the rotating unit and the hydrogen atoms on neighboring units. As shown in \cref{fig:PAQ3_planar}, this leads to significant steric hindrance, effectively preventing free rotation. 

\begin{figure}
\centering
  \includegraphics[width=\linewidth]{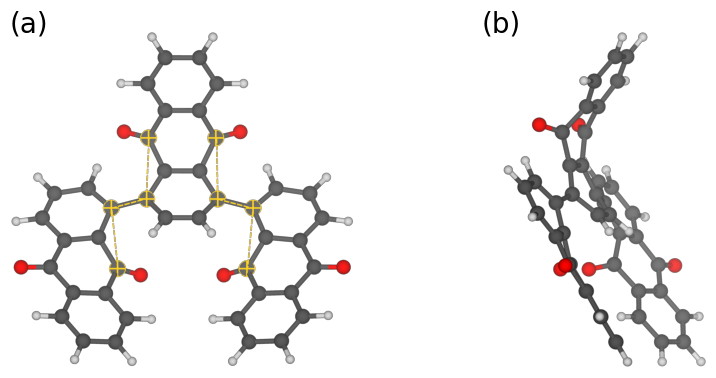}
   \caption{Geometry-optimized structure of P14AQ with three AQ units with dihedral angle constraint, which are marked in yellow in (a), to force an approximation of the extreme structure occurring during rotation of the AQ unit (DFT in ORCA, r2SCAN-3c). (a) and (b) show the same structure from different angles and were both generated with VESTA \cite{mommaVESTA3Threedimensional2011}.} 
   \label{fig:PAQ3_dihedral}
\end{figure}

To approximate the energy needed to rotate the AQ unit, we perform molecular DFT calculations with ORCA of a chain with three AQ units as discussed in Section \ref{sec:methods_rotationAQ}. Herein, we use two dihedral angle constraints between the neighboring AQ units to ${180^{\circ}}$ (marked yellow in \cref{fig:PAQ3_dihedral}a) to force a structure close to the extreme structure for the rotation of the AQ unit. Geometry optimization reveals that the rotating AQ unit bends away from the plane spanned by the neighboring AQ units, with the oxygen atoms displaced in the opposite direction. This optimized structure is shown in \cref{fig:PAQ3_dihedral}. The energy difference between this constrained configuration and the unconstrained, geometry-optimized reference is $\SI{2.4}{eV}$ (level of theory: r2SCAN-3c // pbe0 D3BJ def2-qzvp). As this is not the worst possible structure in terms of steric hindrance, which would occur during rotation due to bending, the actual rotational barrier is expected to be even higher. Consequently, the rotation of an AQ unit through the plane of the neighboring AQ units is not feasible.

We generate all conformers for 8 and 12 AQ units in the unit cell and remove all rotational duplicates as discussed in Section \ref{sec:methods_conformers}. We choose 8 and 12 AQ units due to computational costs and still adequate variation in the orientations of the AQ units.

To investigate the energy difference among P14AQ conformers, we perform DFTB geometry optimizations of the unique conformers with 8 and 12 AQ units in the unit cell. The resulting energy range of the conformers in the 8-AQ unit cell is $\SI{4.2e-2}{eV}$ and $\SI{2.8e-3}{eV}$ in the 12-AQ unit cell (level of theory: DFTB). Due to these small energy differences, all conformers are energetically possible and will be considered for the metal insertion in the following.

\subsection{Energy distributions of structures with one metal atom}
\label{sec:results_ED}
\begin{figure*}
 \centering
 \includegraphics[width=0.7\linewidth]{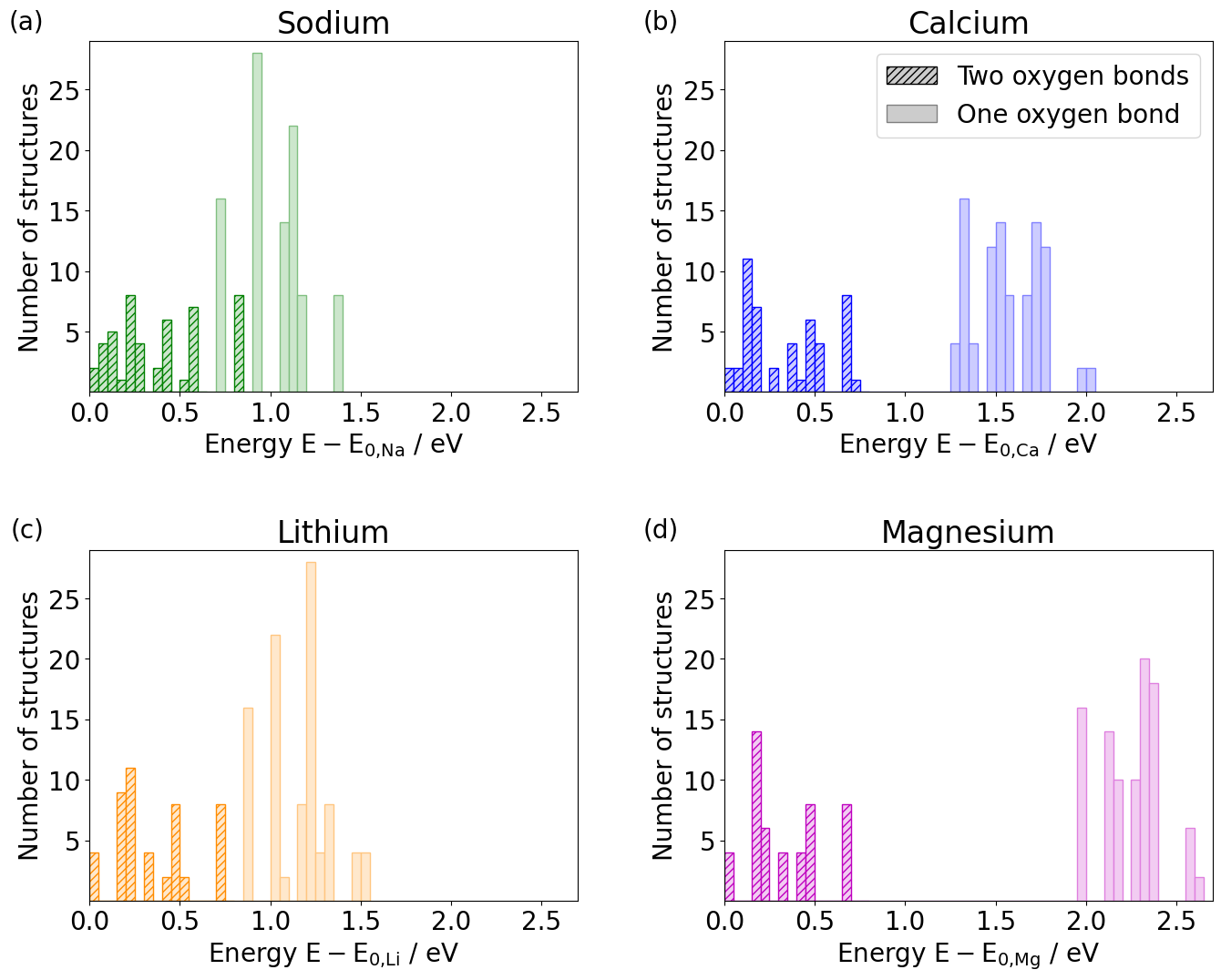}
   \caption{Energy distributions with bin width of $\SI{0.05}{eV}$ of the DFT results of the P14AQ conformers with one single metal atom next to each oxygen side. The structures are multiplied by their number of structures, if no structure after metal insertion would have been removed because of duplicates by rotations. The energy difference is relative to the minimum energy of a structure with that metal. We consider the 8-AQ unit cell.}  
   \label{fig:ED_8_vasp_counts}
\end{figure*}

To uncover the fundamental principles governing metal binding in P14AQ and to model early stages of battery discharge, we investigate the theoretical performance of P14AQ upon insertion of a single metal atom at all possible oxygen sites across the full set of unique conformers. The resulting configurations are ranked by energy, forming an energy distribution that reflects the relative stability of different binding motifs. We compare these energy distributions across monovalent (Li, Na) and divalent (Ca, Mg) metals, revealing distinct coordination preferences and energetic landscapes that underlie their differing electrochemical behavior. 

Since all generated unique conformers of P14AQ are energetically possible, we insert an individual metal atom of Ca, Mg, Na, or Li next to each oxygen atom in all geometry-optimized conformers as discussed in Section \ref{sec:methods_metalinsertion}. With all the resulting structures, we perform DFT geometry optimizations with VASP. Due to the high computational cost of DFT, we use the 8-AQ unit cell since this cell size is sufficient as elaborated in the Supporting Information in Section VI \cite{gausParametrizationBenchmarkDFTB32013, luParametrizationDFTB33OB2015, kubillusParameterizationDFTB3Method2015a,wernerAnalysisOrderingEffects2021, gallmetzerAnthraquinoneItsDerivatives2022} considering DFTB based energy distributions of the 8- and 12-AQ unit cell.

\begin{figure}
\centering
 \includegraphics[width=0.9\linewidth]{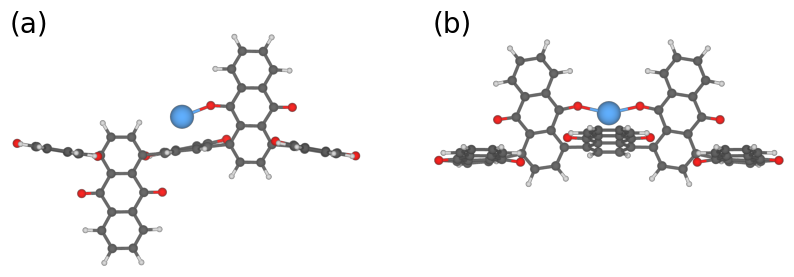}
   \caption{Example of a structure in which the metal atom (blue) is bonded to (a) one or (b) two oxygen atoms. These structures were visualized with VESTA \cite{mommaVESTA3Threedimensional2011}.}   
   \label{fig:BondsMetalOxygen}
\end{figure}

The DFT-based energy distributions of all possible configurations, depicted in \cref{fig:ED_8_vasp_counts}, reveal a fundamental distinction in the energetic landscapes between  monovalent and divalent metals. Ca and Mg exhibit a broad energy range and two regimes separated by a significant energy gap of $\SI{0.52}{eV}$ for Ca and $\SI{1.29}{eV}$ for Mg. As shown in \cref{fig:BondsMetalOxygen}, this is directly governed by the relative orientation of neighboring AQ units. Structures within the lower-energy regime correspond to configurations where the metal atom binds to two oxygen atoms (see \cref{fig:BondsMetalOxygen}b), while higher-energy structures arise when the metal coordinates to only one oxygen atom (see \cref{fig:BondsMetalOxygen}a). 
In contrast, Na and Li exhibit markedly different behavior. Li displays a much smaller gap of $\SI{0.12}{eV}$ between the different bond types and Na shows substantial overlap between the single- and double-bonding regimes. 

Following metal atom insertion into P14AQ, we eliminate rotational duplicates to ensure a non-redundant set of unique configurations. However, in a real battery during discharge, multiple metal atoms are inserted across the structure, with certain sites being more favorable than others. To reflect this, we weighted each structure by the number of positional duplicates in the energy distributions in \cref{fig:BondsMetalOxygen}. We note that this weighting assumes that each conformer occurs with equal probability. The non-weighted energy distribution and a weighted energy distribution with the assumption of uniformly distributed AQ unit orientations in the polymer chain, i.e., weighted by the number of duplicates in each conformer class, are shown and discussed in the Supporting Information in Section VII.

\subsection{Comparison of DFT results with galvanostatic measurements}
\label{sec:comparison_theoretical_experimental}

During battery discharge, metal atoms are inserted into the P14AQ cathode, preferentially occupying the most energetically favorable sites. In the previous section, we computed the energy distributions for single-metal insertion at all possible oxygen sites across the full set of unique conformers. From this single-metal picture, we infer the thermodynamically accessible sites in the fully charged state. If the energy cost of occupying all possible sites remains low, the theoretical capacity is expected to be achievable. Conversely, if certain sites are energetically inaccessible due to high insertion barriers, these positions remain unoccupied, leading to a lower observed capacity. Thus, the single-metal energy distribution provides a framework for understanding the structural and energetic limits of full metal insertion, enabling a direct comparison with experimental capacity data.

As discussed in Section \ref{sec:results_galvanostatic} and depicted in \cref{fig:exp_data_onecycle}, we performed galvanostatic charge and discharge experiments. The measured capacities were $\SI{261}{\milli\amperehour \per \g}$ for Li and $\SI{214}{\milli\amperehour \per \g}$ for Na. Mg and Ca showed lower maximum capacities of $\SI{137}{\milli\amperehour \per \g}$ and $\SI{173}{\milli\amperehour \per \g}$, respectively. Notably, the Na voltage profile displays two distinct voltage plateaus at approximately $\SI{1.5}{\V}$ and $\SI{2.0}{\V}$, while the differential voltage curve for Li shows two small but well-resolved peaks, indicating the formation of two distinct electrochemical phases during cycling.

We compare the experimentally observed capacities to the theoretical capacity of the P14AQ cathode which is $\SI{260}{\milli\amperehour \per \g}$. Counting the structures in \cref{fig:ED_8_vasp_counts}, the metal atom can bind to two oxygen atoms in one third of the structures or $38 \%$ in the unweighted case. Consequently, if a metal atom only binds to two oxygen atoms, the theoretical capacity for monovalent atoms will be around $\SI{87}{\milli\amperehour \per \g}$, and around $\SI{174}{\milli\amperehour \per \g}$ for divalent atoms.

Comparing the experimental results in \cref{fig:exp_data_onecycle} with the calculated energy distributions in \cref{fig:ED_8_vasp_counts}, we find that the energy distribution for Li shows no significant energy gap between single- and double-oxygen binding configurations, suggesting that all possible sites are occupied with Li atoms during cycling. This continuous energy landscape is consistent with the experimental observation of a maximum capacity approaching the theoretical limit. 
Furthermore, the two small but distinct peaks in the differential voltage curve in \cref{fig:exp_data_onecycle}b, indicating two different phases, are consistent with the energy distribution for Li. Structures where Li binds to two oxygen atoms are lower in energy, while those with only one bond are higher in energy but still accessible, explaining the sequential filling of sites and the emergence of two distinct voltage plateaus.

For Na, the charge and discharge curves show a slightly lower capacity compared to Li, with two distinct voltage plateaus. The first plateau ends at approximately $\SI{50}{\milli\amperehour \per \g}$ and the second plateau starts at around $\SI{100}{\milli\amperehour \per \g}$. Since the voltage of a battery cell can be related to the energy of the cathode material with inserted metal atoms \cite{fornariMolecularModelingOrganic2021}, voltage jumps in galvanostatic charge and discharge curves correspond to abrupt changes in the system’s energy, typically indicating phase transitions. This, combined with the energy distribution, indicates a correspondence of the first voltage plateau to structures in which Na binds to two oxygen atoms, and of the second plateau to structures with only one oxygen bond. 

The behavior of divalent metals differs significantly from that of monovalent ions. The experimentally observed maximum capacities for Ca and Mg are close to the theoretical limit expected if these ions bind exclusively to two oxygen atoms, with Ca reaching nearly this value and Mg slightly below it. This lines up with the energy distributions, because Ca and Mg show a large energy gap (Ca: $\SI{0.52}{eV}$, Mg: $\SI{1.29}{eV}$) between structures with one and two bonds to oxygen atoms. This implies that Ca and Mg only bind to sites with two oxygen atoms. Therefore, the comparison of our calculated energy distributions with the galvanostatic measurements elucidates the features observed in the experiments for P14AQ with the different metal atoms.

\section{Conclusion}
In summary, this work demonstrates that the conformational landscape of poly(1,4-anthraquinone) (P14AQ) is not merely a structural detail, it is a fundamental determinant of its electrochemical behavior.
We have demonstrated that it is necessary to consider all accessible conformers of P14AQ because individual anthraquinone (AQ) units are effectively locked in place due to severe steric constraints, preventing rotation through the plane of neighboring units. We developed an automated method to systematically generate all conformers for different numbers of AQ units in the unit cell without rotational duplicates. DFTB geometry optimizations showed that these conformers are energetically similar, indicating that all configurations are thermodynamically accessible and must be included in any accurate modeling of ion binding.

We performed periodic DFT calculations on all conformers of P14AQ with eight AQ units in the unit cell with a single Li, Na, Mg, or Ca atom inserted next to each oxygen site. The resulting energy distributions reveal a critical distinction. While monovalent ions (Li, Na) exhibit continuous energy distributions with minimal gaps between coordination states, divalent ions (Ca, Mg) display a large energy gap between one- and two-oxygen binding configurations. This strongly favors two-site coordination for divalent ions, which is confirmed by the experimental charge–discharge profiles, in which Ca and Mg achieve capacities close to the theoretical limit only if bonding occurs exclusively to two oxygen atoms, explaining their lower gravimetric capacity compared to monovalent ions.

The two distinct voltage plateaus observed in the Na cell directly correspond to the two coordination regimes, two- and one-oxygen binding, while the differential voltage curve for Li shows subtle evidence of this transition.
The strategic placement of two adjacent oxygen atoms for ideal metal binding can be taken into account for the design of future organic cathode materials, particularly those for divalent atoms.

This study is focused on single-metal insertions into the P14AQ unit cell because of computational expenses. We aim to add further metal atoms into the structures in a future study and investigate the binding properties in P14AQ under progressively discharged states.

\section{Author Contributions}

Laura Femmer, Lukas Köbbing, Juliane Heitkämper, and Birger Horstmann conceptualized, developed and performed the theoretical investigation including the conformer generation method, the quantum mechanical calculations and the comparison between the theoretical and experimental results.

Piotr de Silva and Juan Maria García-Lastra supervised the quantum mechanical calculations.

Sibylle Riedel, Devran Cay, and Zhirong Zhao-Karger synthesized the P14AQ and the four electrolyte salts, performed the galvanostatic measurements of the metal-P14AQ cells and the NMR spectroscopy and coordinated the experimental parts. Zhirong Zhao-Karger conceptualized and planned this overall research work.

Alexander J. C. Kuehne performed the GPC and evaluated the experimental polymer analysis (NMR, GPC and MALDI-TOF).

Florin Adler and Birgit Esser performed and evaluated the MALDI-TOF mass spectrometry. Birgit Esser contributed to initial discussions.

\textbf{Laura Femmer:} Conceptualization, Methodology, Software, Validation, Formal analysis, Investigation, Data Curation, Writing - Original Draft, Visualization. 
\textbf{Lukas Köbbing:} Conceptualization, Methodology, Validation, Writing - Review \& Editing, Supervision. 
\textbf{Juliane Heitkämper:} Conceptualization, Methodology, Validation, Writing - Review \& Editing, Supervision. 
\textbf{Sibylle Riedel:} Methodology, Investigation, Data Curation, Writing – Experimental Details, Review \& Editing.
\textbf{Devran Cay:} Investigation.
\textbf{Florin Adler:} Investigation.
\textbf{Birgit Esser:} Resources, Writing - Review \& Editing, Supervision, Funding acquisition.
\textbf{Alexander J. C. Kuehne:} Investigation, Resources, Writing – Polymer analysis, Review \& Editing.
\textbf{Zhirong Zhao-Karger:} Conceptualization, Resources, Writing - Review, Supervision, Funding acquisition.
\textbf{Piotr de Silva:} Methodology, Validation, Writing – Review, Supervision.
\textbf{Juan Maria García-Lastra:} Methodology, Validation, Writing – Review, Supervision.
\textbf{Birger Horstmann:} Conceptualization, Resources, Writing - Review \& Editing, Supervision, Project administration, Funding acquisition.

\section{Acknowledgments}
We thank Markus Lamla from Ulm University for support with the polymer analysis.

This study was funded by the Federal Ministry of Research, Technology and Space of Germany (Bundesministerium für Forschung, Technologie und Raumfahrt, BMFTR) via the projects "CaSino" (03XP0487A and 03XP0487F) and "CaCaO" (03XPB035A and 03XPB035B). Furthermore, we received funding from the German Research Foundation (Deutsche Forschungsgemeinschaft, DFG) under project ID 390874152 (POLiS Cluster of Excellence) and 528773185 (MALDI-TOF spectrometer), and from the European Union under the project EQUALITY (Grant Agreement 101080142). We acknowledge support by the state of Baden-Württemberg through bwHPC and the German Research Foundation (DFG) through grant no INST 40/575-1 FUGG (JUSTUS 2 cluster). This work contributes to the research performed at CELEST (Center for Electrochemical Energy Storage Ulm-Karlsruhe).\\

\textbf{Supporting Information:} Computational and experimental details, additional energy distributions, figures of all unique structures of the 8-AQ unit cell and all measured discharge charge cycles.

\bibliographystyle{apsrev4-2}
\bibliography{bibliography}



\end{document}